\documentclass[epj]{svjour}
\usepackage{graphics}
\DeclareMathAccent{\pol}{\mathord}{letters}{"7E}
\begin{document}
\title{Measurement of the Electric Form Factor of the Neutron \\
at $\mathbf{Q^2 = 0.3\!-\!0.8\; (GeV/c)^2}$}
\author{The A1 Collaboration \\\\
D.I.~Glazier\inst{2}%
\thanks{Comprises part of Ph.D.\ thesis}%
\thanks{\email{d.glazier@physics.gla.ac.uk}}%
, 
M.~Seimetz\inst{1}%
\thanks{Comprises part of doctorate thesis}%
\thanks{Now at: DAPNIA/SPhN, CEA Saclay, 
91191 Gif sur Yvette Cedex, France, \email{mseimetz@cea.fr}}%
, 
J.R.M.~Annand\inst{2}, 
H.~Arenh\"ovel\inst{1}, 
M.~Ases Antelo\inst{1}, 
C.~Ayerbe\inst{1},~%
P.~Bartsch\inst{1}, 
D.~Baumann\inst{1}, 
J.~Bermuth\inst{1}, 
R.~B\"ohm\inst{1}, 
D.~Bosnar\inst{6}, 
M.~Ding\inst{1}, 
M.O.~Distler\inst{1}, 
D.~Elsner\inst{1}\thanks{Now at: Physikalisches Institut, Rheinische Friedrich-Wilhelms-Universit\"at Bonn, Germany}, 
J.~Friedrich\inst{1}, 
S.~Hedicke\inst{1}, 
P.~Jennewein\inst{1}, 
G.~Jover Ma\~nas\inst{1}, 
F.H.~Klein\inst{1}$^{\!\rm e}$, 
F.~Klein\inst{3}, 
M.~Kohl\inst{4}\thanks{Now at: MIT-Bates Linear Accelerator Center, Middleton, USA}, 
K.W.~Krygier\inst{1}, 
K.~Livingston\inst{2}, 
I.J.D.~MacGregor\inst{2}, 
M.~Makek\inst{6}, 
H.~Merkel\inst{1}, 
P.~Merle\inst{1}, 
D.~Middleton\inst{2}, 
U.~M\"uller\inst{1}\thanks{\email{ulm@kph.uni-mainz.de}}, 
R.~Neuhausen\inst{1}, 
L.~Nungesser\inst{1}, 
M.~Ostrick\inst{3}, 
R.~P\'erez Benito\inst{1}, 
J.~Pochodzalla\inst{1}, 
Th.~Pospischil\inst{1}, 
M.~Potokar\inst{5}, 
A.~Reiter\inst{2}, 
G.~Rosner\inst{2}, 
J.~Sanner\inst{1}, 
H.~Schmieden\inst{1}$^{\!\rm e}$, 
A.~S\"ule\inst{1}$^{\!\rm e}$, 
Th.~Walcher\inst{1}, 
D.~Watts\inst{2}, 
M.~Weis\inst{1}
}                     
%
%
\institute{Institut f\"ur Kernphysik, Johannes Gutenberg-Universit\"at Mainz, 
Becherweg 45, 55099 Mainz, Germany \and 
Dept.\ of Physics and Astronomy, University of Glasgow, 
Glasgow G12 8QQ, UK \and 
Physikalisches Institut, Rheinische Friedrich-Wilhelms-Universit\"at Bonn, 
Nussallee 12, 53115 Bonn, Germany \and 
Institut f\"ur Kernphysik, Technische Universit\"at Darmstadt, 
Schlossgartenstra{\ss}e 9, 64289 Darmstadt, Germany \and 
Institute ``Jo\v{ze}f Stefan'', University of Ljubljana, Jamova 39, 1000 Ljubljana, 
Slovenia \and 
Department of Physics, University of Zagreb, Bijeni\v{c}ka c.~32, P.P.~331, 
10002 Zagreb, Croatia}
%
\authorrunning{D.I.~Glazier \emph{et al.}}
\titlerunning{Measurement of the Electric Form Factor of the Neutron}
\date{Received: date / Revised version: date}
%
\abstract{
The electric form factor of the neutron, $G_{E,n}$, has been measured 
at the Mainz Microtron by recoil polarimetry in 
the quasielastic ${\rm D}(\pol{e},e'\pol{n})p$ reaction. 
Three data points have 
been extracted at squared four-momentum transfers 
$Q^2 = 0.3,\; 0.6$ and $0.8\;({\rm GeV}/c)^2$. 
Corrections for nuclear 
binding effects have been applied. 
\PACS{
      {13.40.Gp}{Electromagnetic form factors}   \and 
      {14.20.Dh}{Protons and neutrons}   \and 
      {13.88.+e}{Polarisation in interactions and scattering}
     } 
} 
\maketitle
\section{Introduction}
\label{intro}
The Sachs elastic electromagnetic form factors 
pa\-ram\-e\-trise the nucleon's ability to absorb transferred four-momentum, 
$Q^2$, without excitation or particle emission. They are 
interpreted as the Fourier transforms of the distributions of 
charge and magnetisation inside the nucleon \cite{Ern60,Isg99} 
and may be linked to other observables, such as polarisabilities 
or DIS structure functions, through the framework of Generalised 
Parton Distributions (GPD) \cite{Vdh02}. The elastic form factors 
offer a stringent test of any nucleon structure model. 
Furthermore a precise knowledge of the $Q^2$ dependence of the 
form factors is a prerequisite 
for the interpretation of other electromagnetic 
reactions such as parity-violating $\pol{e}$-$p$ scattering and 
electron-nucleus scattering.

Thus high precision nucleon form factor measurements are necessary, 
but the electric form factor 
of the neutron, $G_{E,n}$, is particularly difficult 
to access, due to its small magnitude and the lack of free 
neutron targets. Using light nuclear targets, ${\rm D}$ and ${}^3{\rm He}$, 
it can be obtained via interference terms where the small $G_{E,n}$ is 
multiplied by the much larger $G_{E,p}$ or $G_{M,n}$ form factors. 
The product $G_{E,n}\, G_{E,p}$ may be accessed in 
\emph{elastic} ${\rm D}(e,e')$ 
scattering. Precision measurements \cite{Pla90} provided $G_{E,n}$ data 
showing a $Q^2$ dependence that can be parametrized by the 
Galster form \cite{Gal71}. 
However these data have large systematic errors, mainly due to 
the model dependent 
uncertainties in unfolding the deuteron wavefunction contribution 
to the cross 
section. Subsequently, reduced model dependence has been achieved from the 
analysis of measurements of the quadrupole form factor, extracted using 
polarisation data \cite{Sch01}. 
Alternatively, uncertainty in model dependent corrections can be 
reduced, and in some cases almost 
eliminated, by measuring asymmetries in double-polarised, quasi-free 
$(e,e'n)$ reactions which yield terms proportional to the 
product $G_{E,n}\, G_{M,n}$. 
Three methods have been used to obtain $G_{E,n}$:
${\rm D}(\pol{e},e'\pol{n})p$ 
\cite{Ede94b,Ost99,Her99,Mad03}, $\pol{\rm D}(\pol{e},e'n)p$, 
\cite{Pas99,Zhu01,War03}, and ${}^3\pol{\rm He}(\pol{e},e')$ 
\cite{Bec99,Ber03}. Bound nucleon corrections, particularly at small 
$Q^2$, are larger and inherently more difficult to calculate for $^3$He. 

The $G_{E,n}$ experiment described here measured the polarisation transfer 
to the neutron in the ${\rm D}(\pol{e},e'\pol{n})p$ reaction. 
The feasibility of the technique was already demonstrated 
at MIT-Bates \cite{Ede94b,Ede94a}. Using the high current, 
high polarisation, 100\%  duty factor electron beam of the Mainz 
Microtron (MAMI), the statistical precision could be improved by an 
order of magnitude. Moreover, the absolute 
calibrations of electron beam polarisation and analysing power 
of the neutron polarimeter 
were avoided by implementing neutron spin precession in an 
appropriate magnetic field \cite{Ost99}. 
Thus this technique, which is explained in sect.~\ref{method}, 
eliminated two major sources of systematic uncertainty. 

Effects of the nuclear binding of the neutron in the deuteron 
were corrected for using the model of Aren\-h\"o\-vel 
\emph{et al.} \cite{Are88}. This model has proven highly successful 
in describing electron 
scattering on the deuteron, so that a small relative error in the 
correction could be assumed. In spite of the corrections' 
sizeable contribution to the final $G_{E,n}$ values, their 
contribution to the systematic uncertainty in $G_{E,n}$ is small.

The following sections provide a description of the experimental 
technique and data analysis, a presentation of the present results 
along with a comparison with previous double-polarisation data, and 
a brief comparison with recent nucleon model calculations.

\section{Experimental method}
\label{method}
\subsection{Principle of measurement}
\begin{figure}[ht]
\begin{center}
\resizebox{\columnwidth}{!}{
\includegraphics{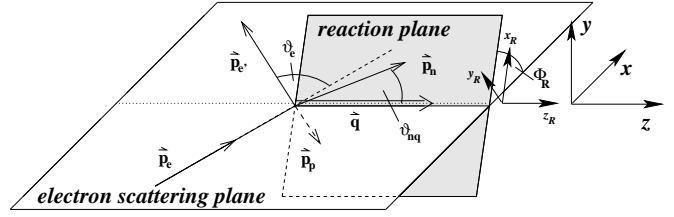}
}
\caption{Definition of the reaction plane, which is spanned by the 
momenta of the outgoing nucleons. The final state neutron is described
in the coordinate system $(x_R,y_R,z_R)$.}
\label{reak}
\end{center}
\end{figure}

In polarised electron-nucleon scattering, $N(\pol{e},e'\pol{N})$, 
the polarisation transfer is favourably expressed in the 
electron scattering plane, which is spanned by the unit vectors 
\begin{equation}\label{frame1} 
\widehat{\vec{z}}=\widehat{\vec{q}},\: \ 
\widehat{\vec{y}}=\frac{\vec{p_{e}}\times  \vec{p_{e}^{\prime}}}{\left|\vec{p_{e}}\times\vec{p_{e}^{\prime}}\right|},\: \ 
\widehat{\vec{x}}=\widehat{\vec{y}}\times \widehat{\vec{z}}\; ,
\end{equation}
as depicted in fig.~\ref{reak}. The ratio of the 
non-vanishing nucleon polarisation components is given by \cite{Akh74,Arn81} 
\begin{equation}\label{polverh}
R_\mathcal{P}=\frac{\mathcal{P}_x}{\mathcal{P}_z} = 
-\frac{1}{\sqrt{\tau + \tau(1+\tau) 
\tan^2 \vartheta_e/2}} \cdot \frac{G_{E,N}}{G_{M,N}}\;,
\end{equation}
where $\tau = Q^2/4M_N^2 c^2$ denotes the dimensionless four 
momentum transfer from the electron to the target 
nucleon, $M_N$ is the nucleon mass, and $\vartheta_e$ is the electron 
scattering angle in the laboratory frame. This formula describes the 
case of a nucleon initially free and at rest. The best 
approximation to this situation is offered by the quasi-free 
${\rm D}(\pol{e},e'\pol{n})p$ reaction. Nuclear binding effects are 
small, provided that the momentum transfer is large compared to the 
nucleons' Fermi-momenta. Corrections for these effects are detailed in 
sect.~\ref{binding}.

Neutron polarimetry exploits the spin-orbit dependence of the strong 
interaction in scattering reactions of polarised neutrons on target 
nuclei. The 
cross section dependence on the azimuthal scattering angle $\Phi_n$ 
is related to the polarisation component transverse to the neutron 
momentum, $\mathcal{P}_t$. It is possible to use 
plastic scintillator material, commonly used in neutron detectors, 
as active neutron scatterer, what allows the reconstruction 
of the interaction vertex \cite{Tad85}. 
If $\mathcal{P}_t \equiv \mathcal{P}_x$ (as for an initially free 
nucleon at rest) the polarised neutron scattering exhibits 
an `up-down' asymmetry, $A$, in the $\Phi_n$ distribution, which may be 
measured with a second scintillator wall. 
It shows a sinusoidal dependence on $\Phi_n$, with an amplitude dependent on 
the analysing power of the first interaction, averaged over the 
detector acceptancies, ${\mathcal A}_{\rm eff}$:
\begin{eqnarray}\label{asymdef}
A & = & \frac{\sqrt{N^+(\Phi_n)N^-(\Phi_n+\pi)} - 
\sqrt{N^+(\Phi_n+\pi)N^-
(\Phi_n)}}{\sqrt{N^+(\Phi_n)N^-(\Phi_n+\pi)} + 
\sqrt{N^+(\Phi_n+\pi)N^-
(\Phi_n)}} \nonumber \\
 & = & {\mathcal A}_{\rm eff} {\mathcal P}_t \sin \Phi_n\;.
\end{eqnarray}
$N^h(\Phi_n)$ and $N^h(\Phi_n+\pi)$ indicate the number of neutrons 
detected in a bin around $\Phi_n$ ($0<\Phi_n<\pi$) and $\Phi_n+\pi$, 
respectively, and with a beam helicity $h=\pm1$. 
This expression exploits the spin flip of the ejected nucleon in 
free $(\pol{e},e'\pol{n})$ scattering under helicity reversal of the 
electron beam and gives a measurement independent of 
the neutron detection efficiency and of the luminosity. 

The need for absolute calibrations of the electron beam 
polarisation, $P_e$, and the effective 
analysing power, ${\mathcal A}_{\rm eff}$, can be circumvented 
by measuring the polarisation ratio 
${\mathcal P}_x/{\mathcal P}_z$ directly. This method was first 
utilised by the 
A3 collaboration at MAMI \cite{Ost99}. In order to become sensitive 
to both the transverse and longitudinal polarisation components, 
the neutron spin is precessed in a magnetic field 
perpendicular to the electron scattering plane. The transverse polarisation 
of the neutron after a precession through angle $\chi$ is given in terms 
of the initial polarisation components (with ${\mathcal P}_y=0$) by 
\begin{equation}\label{prot}
{\mathcal P}_t = 
{\mathcal P}_x \cos \chi - {\mathcal P}_z \sin \chi = 
{\mathcal P}_0 \sin (\chi - \chi_0)\,,
\end{equation}
where ${\mathcal P}_0 = \sqrt{{\mathcal P}_x^2+{\mathcal P}_z^2}$. The 
precession by an angle $\chi_0$ which yields 
a vanishing ${\mathcal P}_t$ and hence zero asymmetry is directly related 
to the form factor ratio through 
\begin{equation}\label{tanchi0}
\tan \chi_0 = R_{\mathcal P} \; .
\end{equation}
As the angle $\chi_0$ is independent of the analysing power and the absolute 
value of the electron beam polarisation, only relative fluctuations of $P_e$ 
have to be monitored when measuring asymmetries for a number of different 
precession angles. 


\subsection{Experimental Setup}
\label{setup}
The ${\rm D}(\pol{e}, e'\pol{n})p$ experiment was carried out 
at the Three Spectrometer Facility of 
the A1 collaboration at the Mainz Microtron, MAMI \cite{Blo98}. The 
polarisation of the incoming electron beam was measured 
regularly with a 
M{\o}ller polarimeter, and was found to be close to 80\%.
The polarised beam, with currents of 
10--15 $\,\mu{\rm A}$, was scattered from a liquid deuterium target 
of $5\,{\rm cm}$ length. 
Spectrometer~A was used to detect the scattered electrons. Its momentum 
resolution $\Delta p/p \le 10^{-4}$ and angular 
resolution $\Delta \varphi, \Delta \vartheta \le 3\,{\rm mrad}$ 
allowed for a precise reconstruction of the virtual photon four-momentum, 
$q$. Data were taken at three central momentum transfers, 
$Q^2 = 0.3,\; 0.6$ and $0.8\;({\rm GeV}/c)^2$. The electron beam 
energy, as well as the central angles of spectrometer~A and of the 
neutron polarimeter, were different in each case. 
Table~\ref{kintable} summarises the kinematics. The pure data 
taking time amounted to 55 days in 2001, and 52 days in 2002.

\begin{table}[ht]
\caption{Kinematics of the A1 $G_{E,n}$ experiment. $E_e'$ 
and $T_n$ are the central energies of the outgoing 
electron and neutron, respectively. The angles $\vartheta_e^c$ and 
$\vartheta_n^c$ are the central angles of the detectors.}
\label{kintable}
\begin{center}
\begin{tabular}{ccccl}
\hline\noalign{\smallskip}
$Q^2/({\rm GeV}/c)^2$ & 0.3 & 0.6 & 0.8 & \\
$E_e/ {\rm MeV}$ & 660 & 855 & 883 & \\
\noalign{\smallskip}\hline\noalign{\smallskip}
$E_e'/ {\rm MeV}$ & 498 & 536 & 454 & 
 \raisebox{-1.5ex}[0ex][-1.5ex]{Spectrometer A} \\
$\vartheta_e^c$ & $57^\circ$ & $70^\circ$ & $90^\circ$ & \\
\noalign{\smallskip}\hline\noalign{\smallskip}
$T_n/ {\rm MeV}$  & 160 & 320 & 427 & Neutron \\
$\vartheta_n^c$ & $47^\circ$ & $37^\circ$ & $27^\circ$ & polarimeter \\
\noalign{\smallskip}\hline
\end{tabular}
\end{center}
\end{table}

\begin{figure*}
\begin{center}
\resizebox{0.9\textwidth}{!}{%
\includegraphics{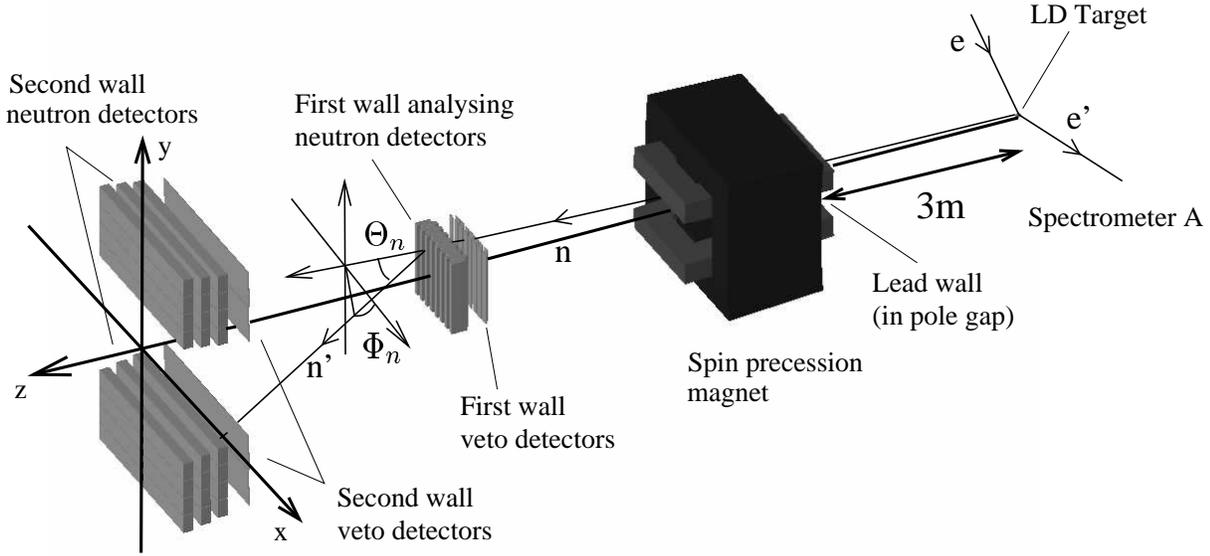}
}
\end{center}
\caption{Three dimensional view of the A1 neutron polarimeter.}
\label{npol1}
\end{figure*}

The neutrons were detected in coincidence with the scattered electrons. 
They passed through a dipole magnet, positioned 3~m from the target, 
the vertical field of which precessed the neutron spin about a vertical 
axis. Analysis scattering of the neutrons took place in a two-layer array of 
plastic scintillators (fig.~\ref{npol1}), each containing 15 
vertically aligned bars 
($5 \times 80 \times 7.5 \;{\rm cm}^3$). This first wall was positioned 
$6\,{\rm m}$ from the target and covered a solid angle of $\sim 17$~msr. 
The light signals from both ends of the scintillator bars were recorded, 
allowing reconstruction (by time difference) of the hit position along 
the length of each bar. Combined with the 5 cm width of the bars this 
resulted in an angular resolution of $0.5^\circ$ FWHM. 
The scintillators of the second wall 
($180 \times 20 \times 10 \;{\rm cm}^3$) were arranged horizontally 
in two blocks above and below the electron scattering plane 
at a distance $3\,{\rm m}$ from the first wall. 
Charged particles were identified by a layer of thin plastic 
scintillators in front of each wall of the polarimeter. The first 
particle-identification (veto) layer consisted of 15 elements of size 
$7.5 \times 81 \times 1 \;{\rm cm}^3$, 
while the rear was constructed from 4 elements per block, of size 
$180 \times 20 \times 1 \;{\rm cm}^3$. 

The rear wall detectors were shielded from direct target view by 
the massive iron yoke of the spin precession magnet. In addition a 
$5\,{\rm cm}$  thick lead wall, located in the gap of the 
spin precession magnet, attenuated mainly low energy electromagnetic 
background from the target region, which was necessary to maintain 
the front-wall single rates at manageable levels ($< 1$~MHz). 
In order to investigate the effect of the lead shielding on the 
polarisation of traversing nucleons, $p(\pol{e},e'\pol{p})$ 
measurements using the focal plane proton polarimeter \cite{Pos02} of 
spectrometer~A have been performed. No indication for 
any degradation of the polarisation has been obtained for protons 
with kinetic energies $T_p \simeq$ 200--350~MeV and shielding 
lengths up to 6~cm \cite{Gro00}. 
A 1~m thick concrete wall, stacked to a height of 5.5~m, 
shielded the 
entire polarimeter from the exit beam line and beam dump.

Full details of the neutron polarimeter 
will be given in a future publication \cite{Sei04}. 

\section{Data analysis}
\subsection{Selection of quasielastic 
$\mathbf{D(\pol{e}, e'\pol{n})p}$ events}
\label{analysis}
\begin{figure}[b] 
\begin{center}
\resizebox{0.24\textwidth}{!}{
\includegraphics{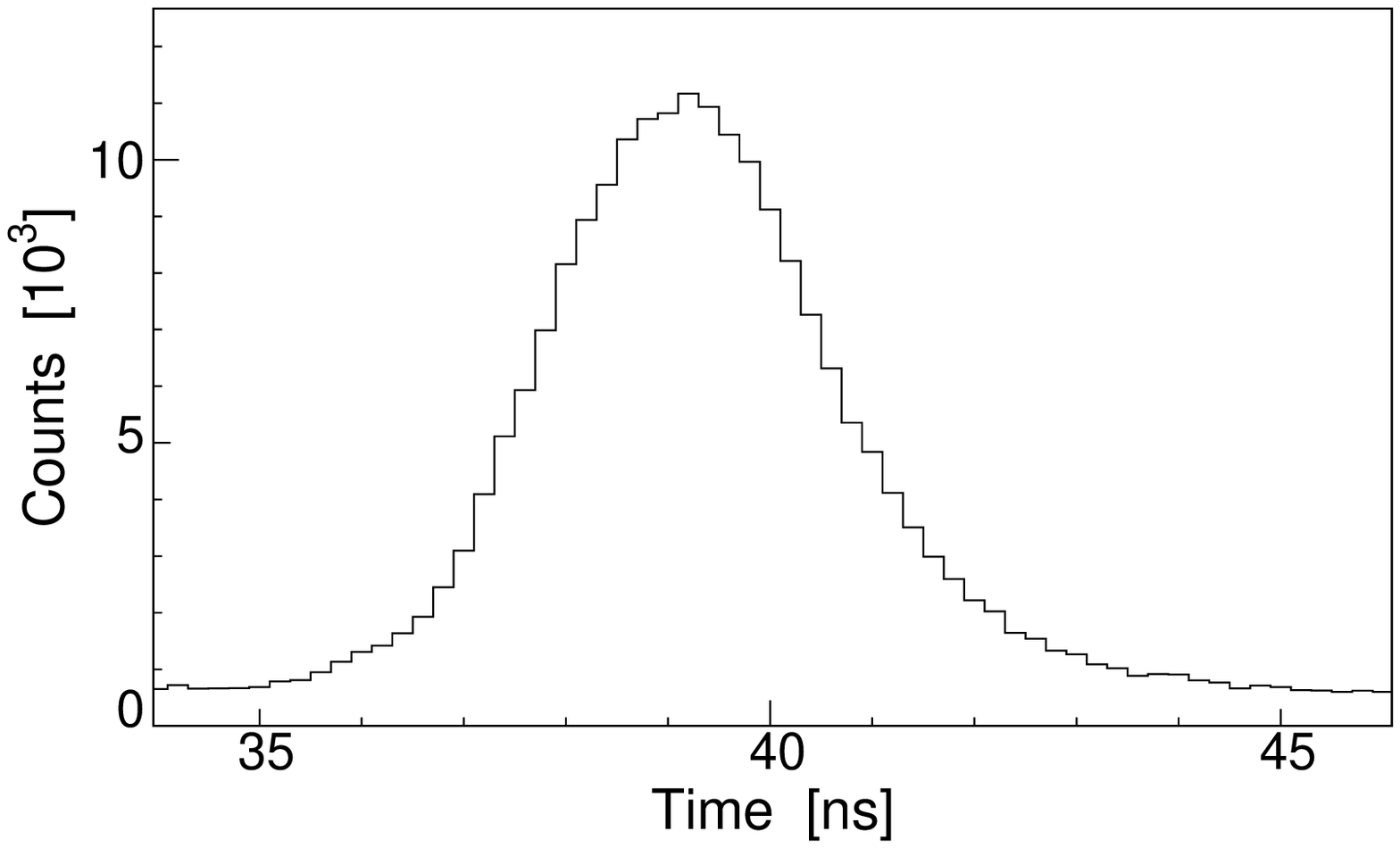}
}
\hfill
\resizebox{0.24\textwidth}{!}{
\includegraphics{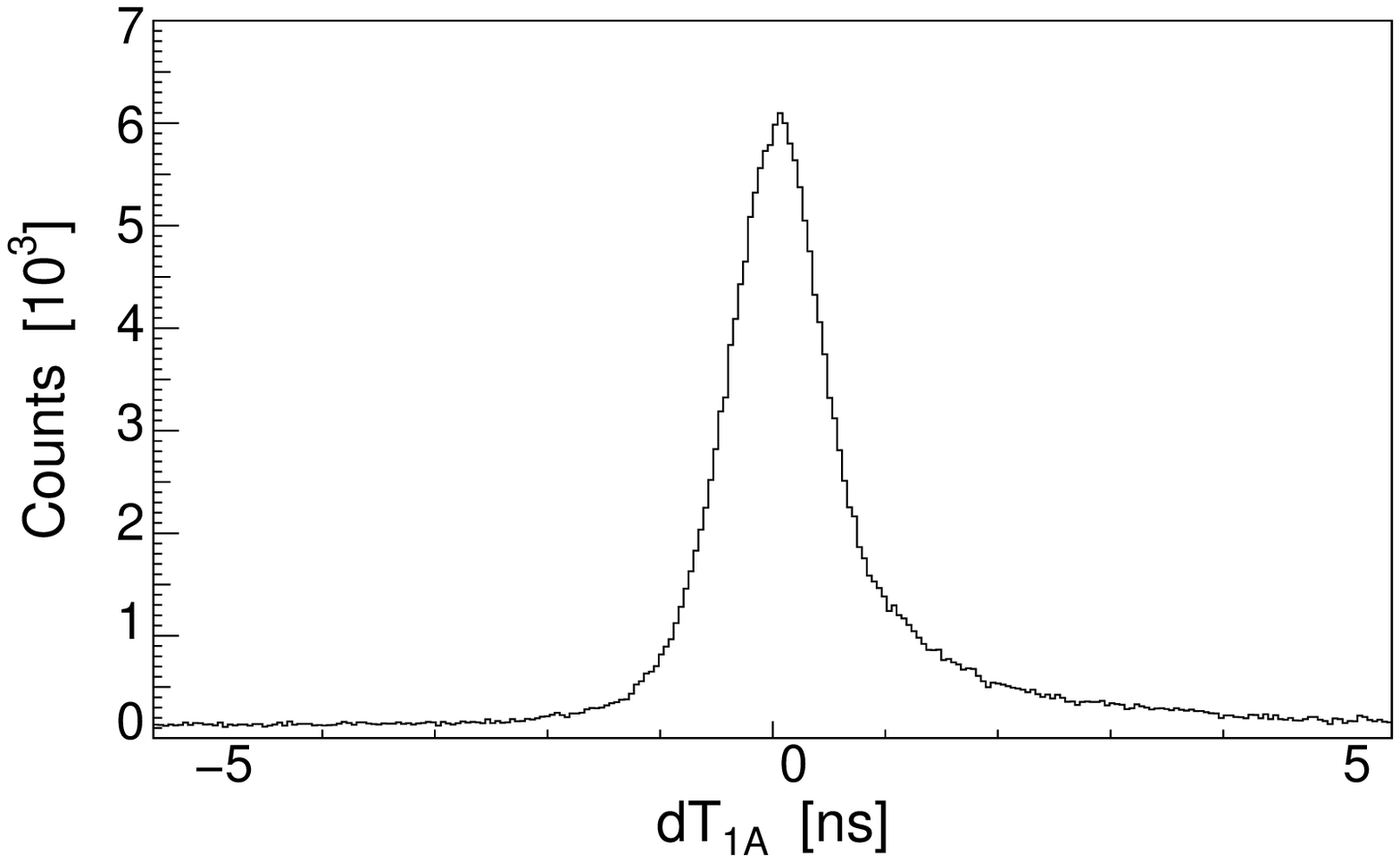}
}
\caption{Left: Measured flight time $T_1$ to the first scintillator wall, 
relative to the electron arrival time in spectrometer A, 
for $Q^2=0.3\;({\rm GeV}/c)^2$. Right: The difference $\Delta T_{1,A}$ 
between $T_1$ and the calculated neutron time-of-flight $T_A$.}
\label{dT1Apic}
\end{center}
\end{figure}
The following observables, measured in the neutron polarimeter and 
spectrometer~A, were relevant for the reconstruction of the deuteron 
breakup reaction: The neutrons' angles, $(\vartheta_n, \varphi_n)$, 
and time-of-flight, $T_{1}$, to the first 
wall and the scattering angles, $(\vartheta_e, \varphi_e)$, 
and energy, $E_e'$, of the electrons. By detecting both particles 
in coincidence the kinematics of the  ${\rm D}(\pol{e}, e'\pol{n})p$ event 
was reconstructed. 
Since deuteron electrodisintegration can be fully described 
with five variables we made use of the sixth for the definition of 
kinematic constraints. For each event the expected neutron 
time-of-flight, denoted by $T_{A}$, as calculated from the set of 
observables $(E_e',\, \vartheta_e,\, 
\varphi_e,\, \vartheta_n,\, \varphi_n)$, was compared to the measured 
time, $T_1$. While the distribution of neutron flight times 
is broad due to Fermi motion and the finite 
acceptances of the spectrometer and of the polarimeter, the variable 
$\Delta T_{1,A}:= T_{1} - T_{A}$ 
has a much narrower distribution which identifies quasielastic events 
with good signal-to-noise ratio (fig.~\ref{dT1Apic}). 

D$(e,e'n)p$ events were identified as follows: 
$\pi^-$ were rejected using threshold \v{C}erenkov information 
of spectrometer~A; 
the $(e,e'n)$ interaction vertex was constrained to be inside 
the target cell; an $e'$-$n$ coincidence was required. 
On the first wall of the polarimeter times-of-flight were 
constrained to be consistent with D$(e,e'n)p$ kinematics. 

\subsection{Neutron scattering in the polarimeter}
\begin{figure}
\begin{center}
\resizebox{0.4\textwidth}{!}{%
 \includegraphics{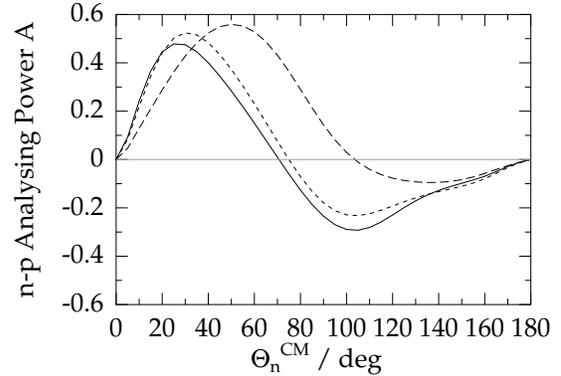}
}
\caption{Analysing power of elastic $n$-$p$ scattering, plotted 
against the neutron scattering angle $\Theta_n^{\rm CM}$ in the 
center-of-mass system for three different neutron energies, 
$T_n = 160$ (long dashed), 319 (short dashed), and $425\,{\rm MeV}$ 
(full curve). Data are taken from SAID \cite{SAID}.}
\label{anapower}
\end{center}
\end{figure}

In addition to the observables listed in sec.~\ref{analysis} the 
scattering angles, $(\Theta_n,\,\Phi_n)$, of particles scattered 
in the first scintillator wall and their time-of-flight, $T_{12}$, 
to the second wall have been measured. 
The analysing power, ${\mathcal A}$, of elastic $n$-$p$ scattering 
is a function of the neutron 
scattering angle and kinetic energy. In fig.~\ref{anapower}, 
${\mathcal A}$ is shown for the three central energies 
associated with our measured $Q^{2}$ values (table ~\ref{kintable}). 
In addition to elastic $n$-$p$ scattering, various inelastic $n$-C 
reactions in the 
mainly ${\rm CH}_2$ scintillator material have an important 
contribution to the observed $(\pol{n},n')$ yield. 
At neutron kinetic energies of a few hundred MeV the quasielastic 
${\rm C}(n,nN)$ channels dominate the cross section and 
have a $\Theta_n$ dependence in analysing power similar to the free $n$-$p$ 
scattering case and therefore contribute positively to the overall 
effective analysing power of the polarimeter. The cross sections and 
analysing powers of these channels are not precisely known, 
and as a result ${\mathcal A}_{\rm eff}$ 
cannot be calculated accurately. Although the spin precession technique 
removes the need for an accurate value of ${\mathcal A}_{\rm eff}$,
the attainable statistical precision in asymmetry measurements depends 
strongly on this quantity. 

Two kinematically distinct regions, corresponding to forward and 
backward $n$-$p$ scattering, are accessible in our polarimeter. 
\begin{enumerate}
\item Neutrons scattered at forward angles transfer only a minor part 
of their kinetic energy to the recoiling protons. The scattered 
neutrons are detected in the rear scintillator wall, while the
10--60~MeV recoil protons are mostly stopped in the front wall. 
In this case the proton energy deposition in 
the front scintillators is correlated with the measured 
time-of-flight of the neutrons between the first and 
second walls. Since the neutrons are detected twice these hits will be 
noted $nn$ events in the following. 
\item The neutron scatters at backward angles and the energetic forward going
recoil proton is detected in the rear scintillator wall. These 
$np$ events have a negative analysing power (fig.~\ref{anapower}) 
with respect to $nn$ events, but since 
the azimuthal angle of detected protons differs from that 
of the neutrons by $180^\circ$ the measured up-down asymmetries have 
the same sign as for the $nn$ case. 
\end{enumerate}
Thus we have two statistically independent and, through the 
charged-particle identification in the rear wall, clearly distinct 
data sets in our analysis. Both sets have comparable statistical 
errors despite a smaller ${\mathcal A}_{\rm eff}$ in the $np$ sample 
due to the 100\% detection efficiency for the protons 
(see sect.~\ref{results}). 
The thin scintillator layers were used to determine charged or 
uncharged hits in the front and rear walls. Charged hits in the front 
wall were rejected while the charged-uncharged decision at the rear 
wall gated the filling of the $np$ or $nn$ samples. The effect of 
proton or neutron misidentification was carefully studied 
(sec.~\ref{results}).

Polarimeter analysis was complicated by the large number of events 
containing multiple hits. Where these occured in adjacent scintillator bars 
we assumed they 
resulted from a single particle and averaged the spatial information
to reconstruct the interaction vertex. 
In the case of hits in separated detector elements 
(\emph{e.g.} in the top and bottom 
 parts of the second wall) separate particles were assumed. In such cases 
all possible combinations of hits in the 
two scintillator walls were analysed as different events and double-counting 
was then corrected for in the background subtraction procedure. The neutron 
time-of-flight spectra showed prominent signals on top of a random background 
and random events were selected from unphysical time-of-flight regions, well 
separated from the signal region.

\begin{figure}
\begin{center}
\resizebox{0.45\textwidth}{!}{%
\includegraphics{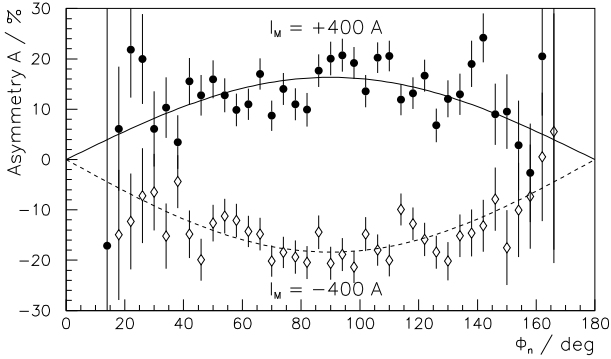} 
}
\caption{The asymmetry $A$ for $nn$ events, plotted against the 
azimuthal scattering angle $\Phi_n$, for two opposite orientations 
of the spin precessing field at magnet currents of 
$I_M = \pm 400\,{\rm A}$ (field integral 1.1~Tm, 
$Q^2 = 0.3\;({\rm GeV}/c)^2$). 
The curves are one-parameter sine fits: $A=A_{0}\sin\Phi_n$.}
\label{asymplot}
\end{center}
\end{figure}

Asymmetries $A$, eq.~(\ref{asymdef}), were generated using the 
$N^{h}(\Phi_n)$ distributions and are shown 
in fig.~\ref{asymplot} for the biggest spin precession angles. 

The zero crossing angle, $\chi_0$, obtained from the $\chi$ 
dependence of the asymmetry 
(eq.s~(\ref{asymdef}) and (\ref{prot})), is independent of the 
absolute value of the 
electron beam polarisation, $P_e$, provided this remains constant. 
The asymmetries at single $\chi$ values depend on $P_e$ at the time of 
measurement. Systematic drifts of the beam polarisation were 
observed only over time scales of several days. Since we changed 
the magnetic field setting every four 
hours the average polarisations of all seven settings are nearly 
equal. The small observed fluctuations in the mean $P_e$ at the 0.2\% 
level, \emph{i.e.}\ within the systematic uncertainty of the M{\o}ller 
polarimeter \cite{Str00}, were factored 
into $A(\chi_i)$ before $\chi_0$ was determined.

\subsection{Correction of nuclear binding effects}
\label{binding}
The relation (\ref{polverh}) between the polarisation 
components of the recoil 
neutron and the Sachs form factors holds exactly for elastic 
scattering of polarised electrons from free nucleons. However in the case of 
quasielastic scattering, the binding and Fermi motion 
of the neutron in the deuteron lead to deviations from the free case. 
The number of independent kinematic 
variables increases from two to five as already mentioned in 
sect.~\ref{analysis}. 
Due to Fermi motion, the recoil neutron 
is not necessarily ejected in the electron 
scattering plane, but rather in a reaction plane, 
which is rotated with respect to the electron scattering plane 
by an angle $\Phi_R$ about the 
direction of momentum transfer, $\mathbf{\hat{q}}$ (fig.~\ref{reak}).
We measured the polarisation transverse to the 
outgoing neutron momentum, and the polarisation components 
in this frame, $\mathcal{P}^R_x$ and $\mathcal{P}^R_z$, 
are related to those in eq.~(\ref{polverh}) through a Wigner rotation. 
The Wigner angle, $\vartheta_W$, is 
closely related to the angle $\vartheta_{nq}$ between the 
direction of $\mathbf{\hat{q}}$ and $\mathbf{\hat{p}_n}$. 
This purely kinematical effect 
cancels in cases where the polarimeter acceptance is perfectly 
$180^{\circ}$-symmetric around $\mathbf{\hat{q}}$. However,
as shown in \cite{Ost99}, small deviations from this ideal 
situation can be corrected for 
by use of a single parameter, $f:=\sin\vartheta_W \cos\Phi_R \ll 1$, 
which is calculated for each event. The polarisation ratios in the electron 
scattering and reaction planes are related via 
\begin{equation}
\frac{\mathcal{P}_x}{\mathcal{P}_z}
 = \frac{\mathcal{P}^R_x}{\mathcal{P}^R_z} + \bar{f} \left(1 + 
\left(\frac{\mathcal{P}^R_x}{\mathcal{P}^R_z}\right)^{\!\!2} \right) + 
\mathcal{O}(\bar{f}^{\,2})\,, 
\label{kincorr}
\end{equation}
using the mean value $\bar{f}$ of the $f$ distribution.
 
Even in quasifree kinematics, 
nuclear binding effects and final state interactions (FSI) 
can lead to a 
reduction of the polarisation ratio compared to
the free case. This is especially significant at 
$Q^2 < 0.3\;({\rm GeV}/c)^2$. 
The effects of FSI, meson exchange currents (MEC), 
and isobar configuration currents (IC) on the polarisation components 
have been calculated in a model by Arenh\"ovel \emph{et al.}\ 
\cite{Are88}. 
In order to incorporate these results into our analysis 
we applied a method similar to that developed by the 
A3 collaboration \cite{Her99}. $R_\mathcal{P}$, 
eq.~(\ref{polverh}), was calculated 
for every accepted event for both the 
free (``Born'' case) and the initially bound (labelled ``FSI'') neutrons. 
The measured zero crossing point of the 
neutron asymmetries was then shifted by 
\begin{equation}
\Delta(\tan\chi_0) = 
\left(\frac{\mathcal{P}_x}{\mathcal{P}_z}\right)_{\rm Born} - 
\left(\frac{\mathcal{P}_x}{\mathcal{P}_z}\right)_{\rm FSI}\;.
\label{fsicorr}
\end{equation}
\begin{figure*}[ht]
\begin{center}
\resizebox{0.45\textwidth}{!}{
\includegraphics{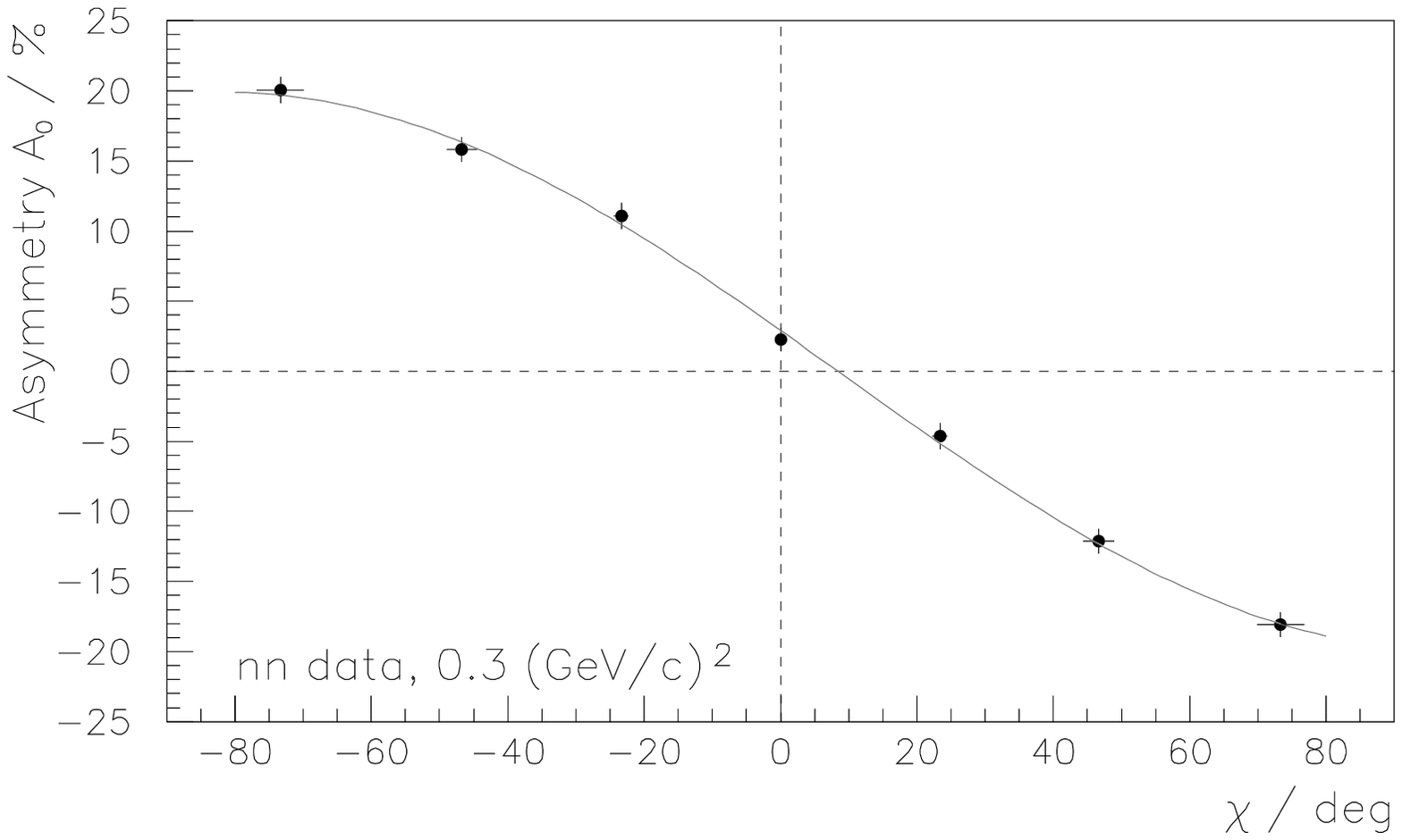}
}
\hfill
\resizebox{0.45\textwidth}{!}{
\includegraphics{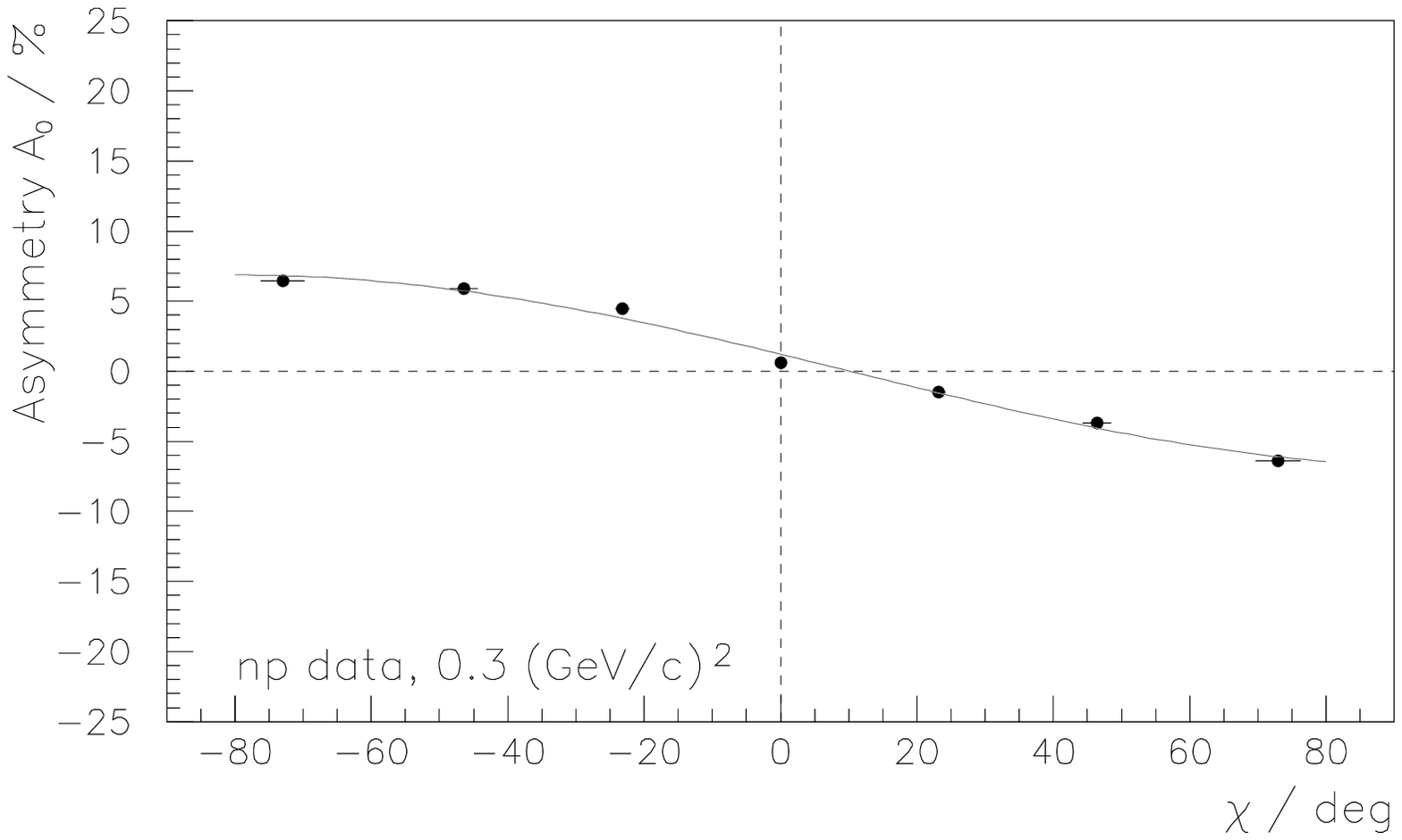}
}

\vspace{1ex}
\resizebox{0.45\textwidth}{!}{
\includegraphics{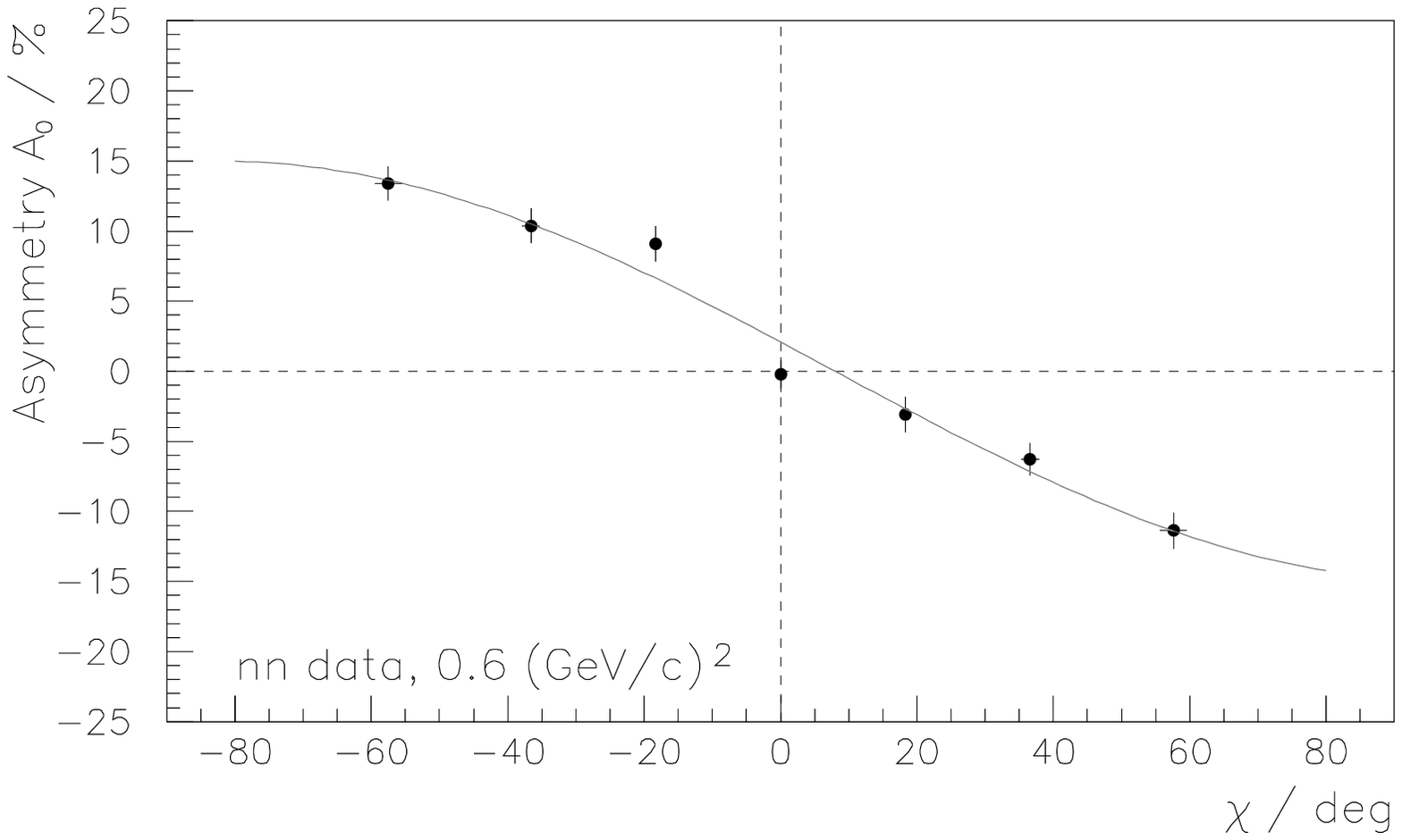}
}
\hfill
\resizebox{0.45\textwidth}{!}{
\includegraphics{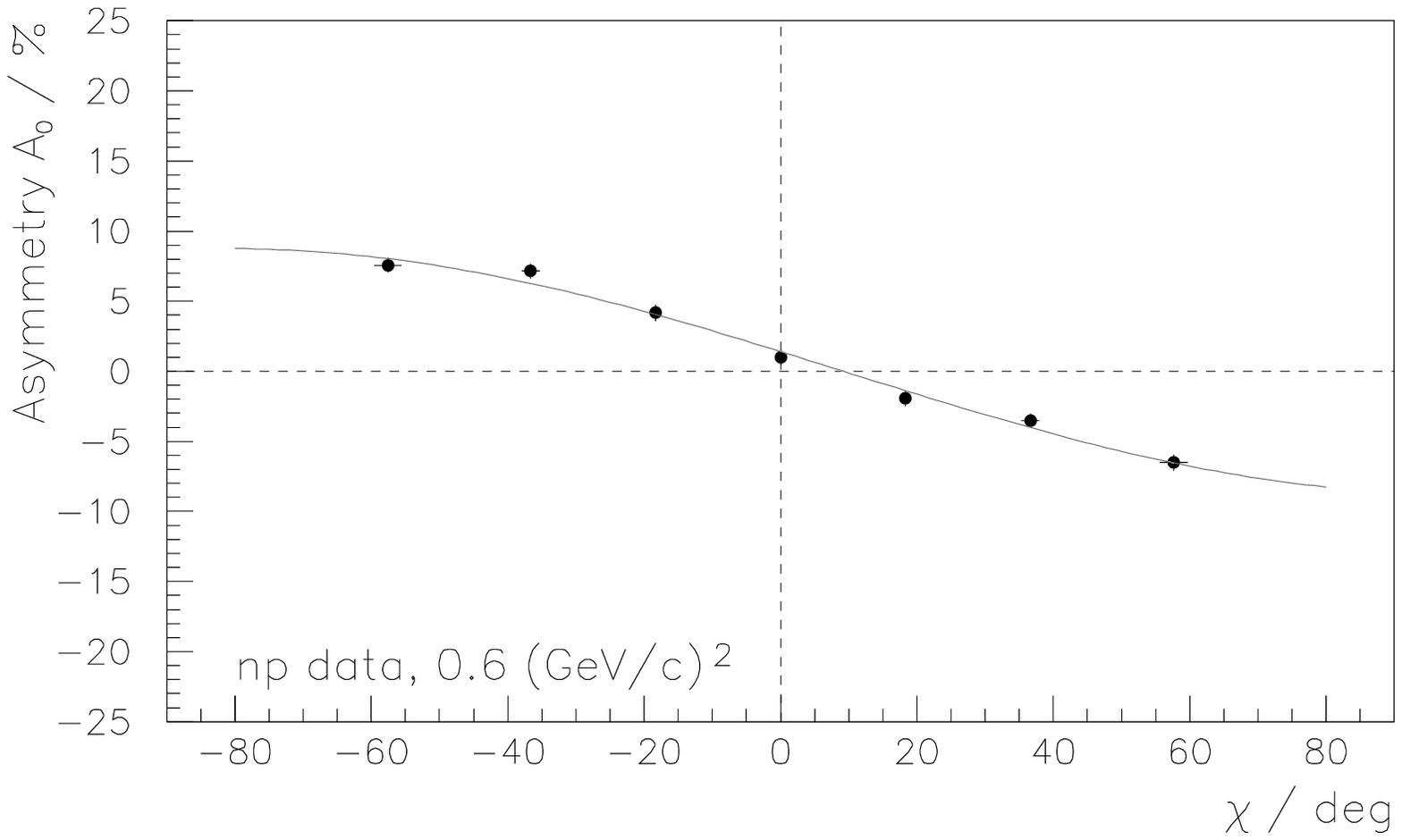}
}

\vspace{1ex}
\resizebox{0.45\textwidth}{!}{
\includegraphics{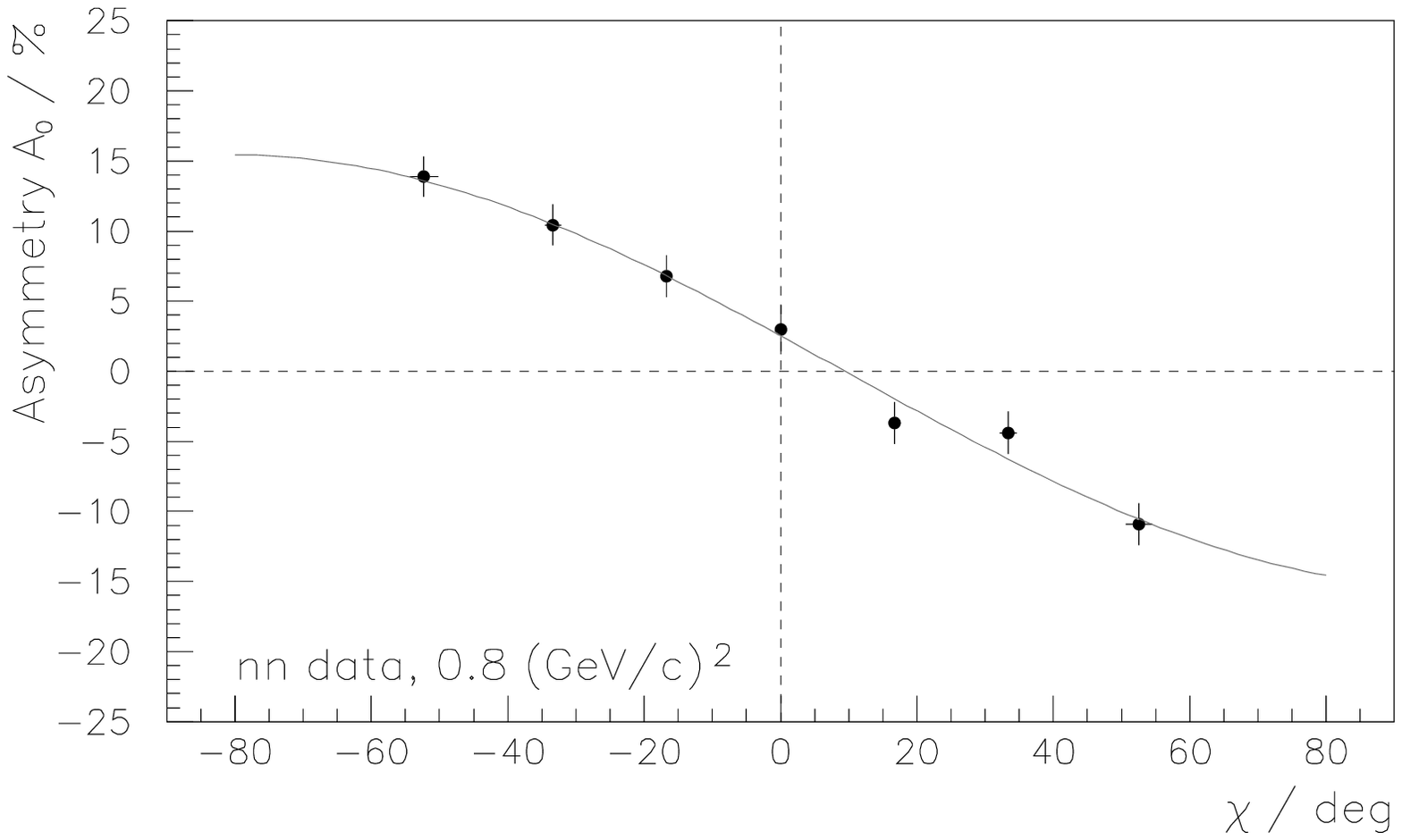}
}
\hfill
\resizebox{0.45\textwidth}{!}{
\includegraphics{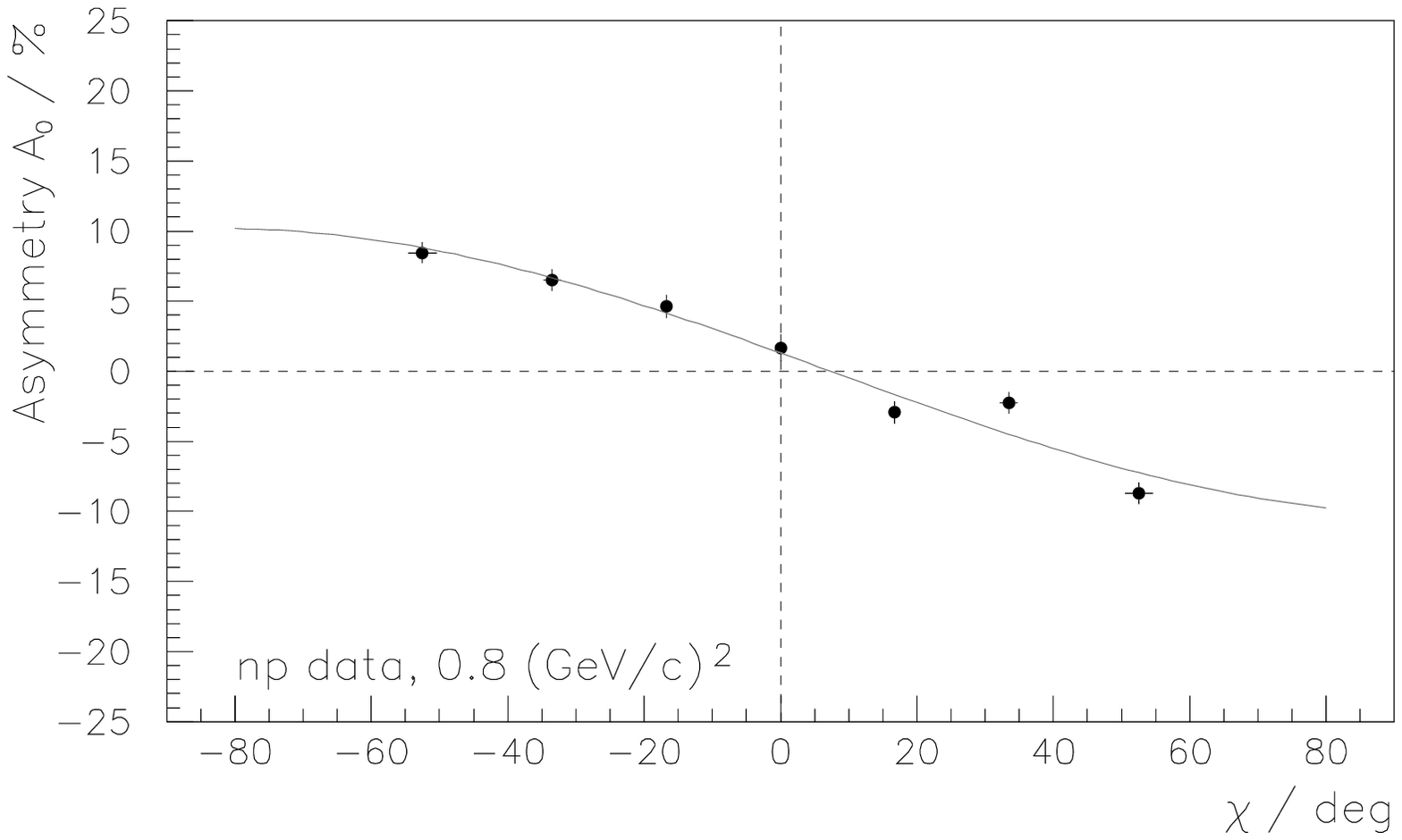}
}
\caption{Fit of the zero crossing point at (from top to bottom) 
$Q^2 = 0.3, \,0.6$ and $0.8\;({\rm GeV}/c)^2$. $nn$ events are shown on the 
left, $np$ events on the right hand side. Horizontal error bars represent 
the uncertainties in $\chi$ (see text).}
\label{asymplots}
\end{center}
\end{figure*}
The effect of the weak dependence of the corrections on the input value 
for $G_{E,n}$ was estimated by varying this value in the model 
over a reasonable range.

\section{Results}
\subsection{Measured data points}
\label{results}
In fig.~\ref{asymplots} we show the measured $\chi$ dependence 
of the azimuthal asymmetries $A_0$ for the momentum transfers 
$Q^2 = 0.3$, 0.6 and $0.8\;({\rm GeV}/c)^2$ and for the 
$nn$ and $np$ data samples. The zero crossing points $\chi_0$ were 
obtained via two-pa\-ra\-me\-ter sine fits: 
$A_0=A_0^0 \sin(\chi-\chi_0)$. The asymmetries were larger in the 
$nn$ case due to the larger analysing power of this channel. 
However, the statistical errors on the extracted $\chi_0$ values 
are of similar size for the $nn$ and $np$ samples 
due to the larger number of detected $np$ events. The results of the fits 
as well as the acceptance and nuclear binding corrections are shown 
in table~\ref{genresults}.

The extraction of $G_{E,n}$ through eq.~(\ref{polverh}) 
requires the knowledge of the magnetic form factor at the given 
$Q^2$ and of the kinematic factor 
\begin{equation}
F:=- {[\tau+\tau(1+\tau) \tan^2 \vartheta_e/2]}^{-1/2}\;. 
\end{equation}
The former was taken from 
a recent parametrisation \cite{Kub02} which gives a good 
representation of $G_{M,n}$ throughout our $Q^2$ range. It is based 
on data obtained from five experiments and 
quotes a relative error of $\Delta G_{M,n}/G_{M,n} \simeq 1.1\%$. 
The factor $G_{M,n} \cdot F$ was calculated for each event and 
then averaged over the acceptance.

Our  final $G_{E,n}$ values were obtained from the mean values for the 
$nn$ and $np$ samples taken at the same $Q^2$. 
They are summarised in table~\ref{genresults}.

\begin{table*}[ht] 
\caption{Results for the $nn$ and $np$ data samples at the three 
$Q^2$ values (the ranges in $Q^2$ represent the experimental 
acceptance): 
Numbers of accepted events, $N_n$, asymmetry amplitudes, $A_0^0$, 
and the results for $G_{E,n}$ extracted from this experiment. Values 
for $G_{E,n}$ are given for an evaluation of the 
uncorrected asymmetries, $G_{E,n}^{\,\rm uncorr.}$, for an analysis 
accounting for the kinematic corrections of nuclear binding effects 
via eq.~(\ref{kincorr}), $G_{E,n}^{\,\rm kin.\ corr.}$, and for the 
final results, $G_{E,n}^{\,\rm full\ corr.}$, where also the FSI 
corrections, eq.~(\ref{fsicorr}), have been taken into account. 
The last column contains the combined results of the $nn$ and $np$ samples.}
\label{genresults}
\begin{center}
\begin{tabular}{ccrrcccc}
\hline\noalign{\smallskip}
$Q^2 / ({\rm GeV}/c)^2$ & sample & \multicolumn{1}{c}{$N_n$} & 
\multicolumn{1}{c}{$A_0^0 \cdot 100$} & $G_{E,n}^{\,\rm uncorr.}$ & 
 $G_{E,n}^{\,\rm kin.\ corr.}$ & 
 $G_{E,n}^{\,\rm full\ corr.} \pm \Delta G_{E,n}^{\,\rm stat}$ & 
 $G_{E,n} \pm \Delta G_{E,n}^{\,\rm stat} \pm \Delta G_{E,n}^{\,\rm syst}$ \\
\noalign{\smallskip}\hline\noalign{\smallskip}
\rule[-0.5ex]{0ex}{3.0ex} $0.30 \pm 0.02$ & $nn$ & 114000 & $19.9 \pm 0.5$ & 
 0.0458 & 0.0447 & $0.0520 \pm 0.0077$ & \\
\rule[-1.5ex]{0ex}{4.0ex} $0.30 \pm 0.02$ & $np$ & 570000 & $ 6.9 \pm 0.5$ & 
 0.0555 & 0.0524 & $0.0607 \pm 0.0100$ & 
 \raisebox{1.5ex}[0ex][1.5ex]{$0.0552 \pm 0.0061 {}^{+0.0018}_{-0.0011}$} \\ 
\rule[-0.5ex]{0ex}{3.0ex} $0.59 \pm 0.03$ & $nn$ &  55000 & $15.0 \pm 0.8$ & 
 0.0413 & 0.0408 & $0.0437 \pm 0.0116$ & \\
\rule[-1.5ex]{0ex}{4.0ex} $0.59 \pm 0.03$ & $np$ & 316000 & $ 8.8 \pm 0.4$ & 
 0.0475 & 0.0469 & $0.0500 \pm 0.0088$ & 
 \raisebox{1.5ex}[0ex][1.5ex]{$0.0477 \pm 0.0070 {}^{+0.0019}_{-0.0008}$} \\ 
\rule[-0.5ex]{0ex}{3.0ex} $0.79 \pm 0.03$ & $nn$ &  86000 & $15.4 \pm 1.0$ & 
 0.0526 & 0.0519 & $0.0545 \pm 0.0146$ & \\
\rule[-1.5ex]{0ex}{4.0ex} $0.79 \pm 0.03$ & $np$ & 204000 & $10.2 \pm 0.5$ & 
 0.0402 & 0.0396 & $0.0420 \pm 0.0116$ & 
 \raisebox{1.5ex}[0ex][1.5ex]{$0.0468 \pm 0.0090 {}^{+0.0025}_{-0.0010}$} \\ 
\noalign{\smallskip}\hline
\end{tabular}
\end{center}
\end{table*}

The systematic errors are not symmetric since 
background events due to misidentified protons may cause 
asymmetries with signs opposite to the ones expected for neutrons. 
Protons may be misidentified in the first scintillator wall for two 
reasons. First of all, even though charged particles are eliminated 
with the use of veto detectors and additional offline hit pattern 
conditions this filtering may not be 100\% efficient. Secondly, 
charge-exchange reactions in the lead shielding may result in 
$p$-$n$ conversion. 
In order to estimate the magnitude of these effects data 
taken with an ${\rm LH}_2$ target have been analysed applying the 
same event selection conditions as for the ${\rm LD}_2$ data, and 
possible asymmetries caused by the false-identified neutron events have 
been estimated conservatively assuming that 
all misidentified protons were polarised according to 
eq.~(\ref{polverh}). Due to limited statistics a direct extraction 
of false asymmetries from the ${\rm LH}_2$ data was not 
feasible. Therefore the ``worst-case'' estimate is considered as 
contribution to the error in the determination of $G_{E,n}$, rather 
than as a correction term. It is given in table~\ref{systable} along with 
the other sources of systematic uncertainty. 

As mentioned above, small 
fluctuations of the beam polarisation have been corrected for in the 
data analysis. Uncertainties in the neutron precession angles, 
resulting from field inhomogeneities, the absolute calibration of 
the field integrals, and the finite acceptance in 
neutron velocities have been included as ordinate errors 
in the fit procedure of the zero crossing points. The influence of 
radiative corrections on the extracted form factor ratio is estimated 
on basis of the calculation of Afanasev \emph{et al.} \cite{Afa01} 
to be well below the one percent level.

\begin{figure*}[ht]
\begin{center}
\resizebox{0.7\textwidth}{!}{\includegraphics{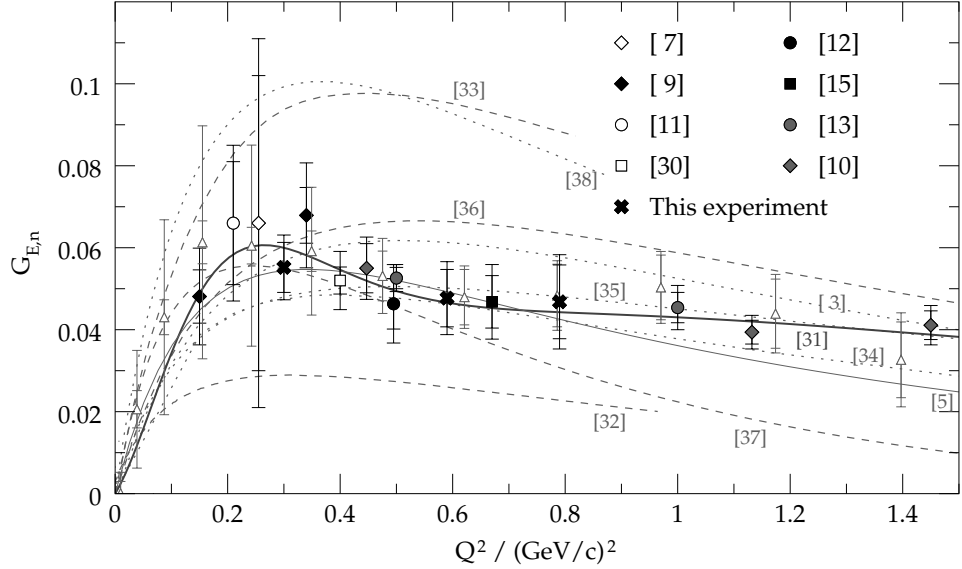}}
\caption{$G_{E,n}$ from double-polarisation experiments. 
Polarisation-transfer measurements on the deuteron 
\cite{Ede94b,Her99,Mad03} are marked with diamonds, 
experiments using polarised Deuterium \cite{Pas99,Zhu01,War03} or 
${}^3\pol{\rm He}$ \cite{Ber03,Gol01} targets are shown as circles and 
squares, respectively. Open triangles 
refer to the analysis \cite{Sch01} of unpolarised data. The thin 
full curve represents the original Galster parametrisation 
\cite{Gal71}, the thick line represents the ``pion-cloud'' 
parametrisation \cite{Fri03} (see text). The dashed and 
dotted lines are discussed in the text.}
\label{gen_newresults}
\end{center}
\end{figure*}

\begin{table}[ht]
\caption{Systematic errors in the extraction of $G_{E,n}$. The 
quoted ranges cover the three $Q^2$ values.}
\label{systable}
\begin{center}
\begin{tabular}{lcrc}
\hline\noalign{\smallskip}
 Error source & \multicolumn{3}{c}{$\Delta G_{E,n} / G_{E,n}$} \\
\noalign{\smallskip}\hline\noalign{\smallskip}
 $G_{M,n}$                          & &  1.1\% & \\
 $F \cdot G_{M,n}$                  & &  0.9\% & \\
 Remaining proton background        & & 2.5--5.0\% & \\
 FSI corrections                    & & 0.7--1.4\% & \\
\noalign{\smallskip}\hline\noalign{\smallskip}
 total                              & & 3.0--5.2\% & \\
\noalign{\smallskip}\hline
\end{tabular}
\end{center}
\end{table}

Our results are shown as crosses in fig.~\ref{gen_newresults} together 
with published $G_{E,n}$ values from other double-polarisation experiments. 
The new data points at $Q^2 = 0.3,\; 0.6$ and $0.8\;({\rm GeV}/c)^2$ 
agree well with those of the previous measurements. 
Some $G_{E,n}$ data points have been 
superseded and therefore are not shown here. 
This holds \emph{e.g.} for the pilot experiment of the A3 collaboration 
\cite{Mey94}, which was later repeated with much better 
statistics \cite{Bec99}. FSI corrections, which are sizeable 
for ${}^3\pol{\rm He}$ at low $Q^2$, were applied to these results \emph{a 
posteriori} in an independent publication \cite{Gol01}, which, in 
turn, ignored the systematic experimental errors. The respective 
A3 data point included in fig.~\ref{gen_newresults} is the 
FSI-corrected result from \cite{Gol01}, but with the systematic 
errors from the original paper \cite{Bec99} added, which gives a 
fair representation of the results of that experiment.

\subsection{Phenomenological Fits}
The thin full line in fig.~\ref{gen_newresults} shows the simple 
$G_{E,n}$ para\-metrisation given by Galster \emph{et al.}\ 
in 1971 \cite{Gal71}, 
\begin{equation}
G_{E,n}(Q^2) = - \frac{\mu_n \tau}{1 + p \tau} G_D(Q^2)\,
\end{equation}
with the usual dipole form factor $G_D$. It was purely 
phenomenological and contained only one 
free parameter, which was determined at that time from data 
up to $1\;({\rm GeV}/c)^2$ to be $p = 5.6$. This fit is still in 
reasonable agreement with the recent double-polarisation data, but 
considering the large uncertainties in the data available in 1971, 
this has to be regarded as coincidental. 

Recently the low 
$Q^2$ dependence of the Sachs form factors has been 
interpreted \cite{Fri03} as direct evidence for a pion cloud surrounding 
the bare nucleon. The thick, full curve in fig.~\ref{gen_newresults} 
represents the phenomenological parametrisation of ref.~\cite{Fri03}, 
showing a bump, characteristic of the pion cloud, at 
$Q^2 \simeq 0.25\;({\rm GeV}/c)^2$. It has the form 
\begin{equation}\label{fw1}
G_N(Q^2) = G_s(Q^2) + a_b \cdot Q^2 G_b(Q^2)\;,
\end{equation}
with a smooth part 
\begin{equation}\label{fw2}
G_s(Q^2) = \frac{a_{10}}{(1+Q^2/a_{11})^2} + \frac{a_{20}}{(1+Q^2/a_{21})^2} 
\end{equation}
and a bump 
\begin{equation}\label{fw3}
G_b(Q^2) = e^{-\frac{1}{2} (\frac{Q-Q_b}{\sigma_b})^2} + 
e^{-\frac{1}{2} (\frac{Q+Q_b}{\sigma_b})^2}\;,
\end{equation}
where $Q = \sqrt{Q^2}$. 
A fit to the current data yields the parameters 
$a_{10} = 1.2974$, $a_{20} = -a_{10}$ (for normalisation), 
$a_{11} =  1.73010\;({\rm GeV}/c)^2$ (fixed in the fit), 
$a_{21} =  1.54479\;({\rm GeV}/c)^2$, 
$a_b =  0.19426$ $({\rm GeV}/c)^{-2}$, $Q_b =  0.34210$ ${\rm GeV}/c$, and 
$\sigma_b =  0.16758\;{\rm GeV}/c$. 
The slow fall of $G_{E,n}$ at higher $Q^2$ is 
accommodated by the smooth part of the ansatz, eq.~(\ref{fw2}).

\section{Nucleon Models}
\label{discussion}
The elastic nucleon form factors present a significant test for 
nucleon models and most recent calculations aim to reproduce all 
four electromagnetic form factors with one set of adjustable parameters. 
Thus a full discussion of the predictive power of various models 
requires comparison with the complete elastic form factor data set. 
Such a comparison is beyond the scope of the present work and we 
confine our discussion to some recent calculations of $G_{E,n}$. The 
curves showing the predicted $G_{E,n}$ values 
(fig.~\ref{gen_newresults}) are labeled with the present reference of 
the particular calculation.

\begin{enumerate}
\item Nucleon models built upon basic assumptions and with a 
small number of free 
parameters include the semibosonized SU(3) 
Nambu--Jona-Lasinio model \cite{Kim96} and relativistic constituent 
quark models \cite{Mer02}. 
They are, in general, not able to provide a satisfactory description of 
all form factors, a fact which underlines the lack of understanding of the 
structure of the nucleon.
\item Generalised Parton Distributions (GPD) are considered to 
represent the momentum distributions of the constituents of the 
nucleon and the elastic form factors represent moments of particular GPDs. 
With a simple Regge ansatz for the $Q^2$-dependence of the 
GPDs $H$ and $E$ it is possible to reproduce experimental form factor 
data over a large range of $Q^2$ \cite{Vdh02}.
\item A recent vector dominance model \cite{Lom02} gives a reasonably 
good description of the four elastic proton and neutron form factors. 
\item A dispersion-theoretical analysis \cite{Ham04} reproduces the 
trend of the electromagnetic form factors over a wide range of 
momentum transfers. At small $Q^2$ however it suggests values for 
$G_{E,n}$ which are significantly smaller than the measured data 
points. 
\item The diquark-quark model of ref.~\cite{Ma02} describes the 
measured $G_{E,n}$ at low $Q^2$, it misses however the trend of the 
data for $Q^2 > 0.4\;({\rm GeV}/c)^2$. 
\item The chiral soliton model of ref.~\cite{Hol02} falls far below 
the $G_{E,n}$ data for $Q^2 > 0.6\;({\rm GeV}/c)^2$. In addition, 
some of its parameters are clearly at variance with 
experiment, such as the anomalous magnetic moments of the proton and 
neutron. 
\item Recent lattice QCD calculations in quenched approximation 
\cite{Ash04} qualitatively reproduce the trend of the data for 
the four electromagnetic 
nucleon form factors. Quantitative discrepancies, however, amount to 
a factor 2 in case of $G_{E,n}$.
\end{enumerate}

\section{Conclusion}
\label{conclusion}
The electric form factor of the neutron, $G_{E,n}$, has been measured 
at four-momentum transfers 0.3, 0.6, and 
$0.8$ $({\rm GeV}/c)^2$ in a double-polarisation experiment 
using a polarised electron beam and a final-state neutron 
polarimeter in the reaction 
${\rm D}(\pol{e}, e'\pol{n})p$. The ratio of transverse to longitudinal 
neutron polarisation components was measured by precession of the neutron 
spin in a magnetic field, which provided a cancellation 
of several systematic uncertainties. 
Nuclear binding effects have been corrected for using a model which gives an 
excellent account of a broad range of electron-deuteron reactions. The 
present experimental results are in good agreement with all other 
$G_{E,n}$ double-polarisation measurements.

\section*{Acknowledgements}
We are highly indebted to the MAMI accelerator group and 
the MAMI polarised beam group for their outstanding performance. 
This work was supported by the Sonderforschungsbereich 
443 of the Deutsche Forschungsgemeinschaft and by the UK Engineering 
and Physical Sciences Research Council. 



\begin{thebibliography}{99}
%
%

\bibitem{Ern60} F.J.~Ernst, R.G.~Sachs, K.C.~Wali, 
Phys.~Rev. \textbf{119}, 1105--1114 (1960).
\bibitem{Isg99} N.~Isgur, Phys.~Rev.~Lett.\ \textbf{83}, 272--275 (1999).
\bibitem{Vdh02} M.~Vanderhaeghen, \textit{Exclusive Processes at high 
momentum transfer}, (Eds. A.~Radyushkin, P.~Stoler, World Scientific, 
Singapore 2002), 51--56.
\bibitem{Pla90} S.~Platchkov \emph{et al.}, 
Nucl.~Phys.~A \textbf{510}, 740--758 (1990).
\bibitem{Gal71} S.~Galster \emph{et al.}, 
Nucl.~Phys.~B \textbf{32}, 221--237 (1971).
\bibitem{Sch01} R.~Schiavilla, I.~Sick, Phys.~Rev.~C \textbf{64}, 
041002 (2001).
\bibitem{Ede94b} T.~Eden \emph{et al.}, 
Phys.~Rev. C \textbf{50}, 1749--1753 (1994).
\bibitem{Ost99} M.~Ostrick \emph{et al.}, 
Phys.~Rev.~Lett.\ \textbf{83}, 276--279 (1999).
\bibitem{Her99} C.~Herberg \emph{et al.}, 
Eur.~Phys.~J.~A \textbf{5}, 131--135 (1999).
\bibitem{Mad03} R.~Madey \emph{et al.}, Phys.~Rev.~Lett.\ 
\textbf{91}, 122002 (2003).
\bibitem{Pas99} I.~Passchier \emph{et al.}, Phys.~Rev.~Lett.\ 
\textbf{82}, 4988--4991 (1999).
\bibitem{Zhu01} H.~Zhu \emph{et al.}, 
Phys.~Rev.~Lett.\ \textbf{87}, 081801 (2001).
\bibitem{War03} G.~Warren \emph{et al.}, 
Phys.~Rev.~Lett.\ \textbf{92}, 042301 (2004).
\bibitem{Bec99} J.~Becker \emph{et al.}, Eur.~Phys.~J.~A \textbf{6}, 
329--344 (1999).
\bibitem{Ber03} J.~Bermuth \emph{et al.}, 
Phys.~Lett.~B \textbf{564}, 199--204 (2003).
\bibitem{Ede94a} T.~Eden \emph{et al.}, 
Nucl.~Instrum.~Meth.~A \textbf{338}, 432--441 (1994).
\bibitem{Are88} H.~Arenh\"ovel, W.~Leidemann, E.L.~Tomusiak, 
Z.~Phys.~A \textbf{331}, 123--138 (1988).
\bibitem{Akh74} A.I.~Akhiezer, M.P.~Rekalo, Sov.~J.~Nucl.~Phys. \textbf{4}, 
277--287 (1974).
\bibitem{Arn81} R.G.~Arnold, C.E.~Carlson, F.~Gross, 
Phys.~Rev.~C \textbf{23}, 363--374 (1981).
\bibitem{Tad85} T.N.~Taddeucci \emph{et al.}, Nucl.~Instrum.~Meth.~A 
\textbf{241}, 448--460 (1985).
\bibitem{Blo98} K.I.~Blomqvist \emph{et al.}, 
Nucl.~Instrum.~Meth.~A \textbf{403}, 263--301 (1998).
\bibitem{Pos02} Th.~Pospischil \emph{et al.}, Nucl.~Instrum.~Meth.~A 
\textbf{483}, 713--725 (2002).
\bibitem{Gro00} S.O.~Gr\"ozinger, diploma thesis, 
Institut f\"ur Kernphysik, Mainz, Germany 2001.
\bibitem{Sei04} M.~Seimetz \emph{et al.}, \emph{A Neutron Polarimeter 
for the Three Spectrometer Setup at MAMI}, to be submitted to 
Nucl.~Instrum.~Meth.~A.
\bibitem{SAID} R.A.~Arndt, I.I.~Strakovsky, R.L.~Workman, 
Phys.~Rev.~C \textbf{62} (2000), 034005; \verb+http://gwdac.phys.gwu.edu+ .
\bibitem{Str00} O.~von Str\"ahle, diploma thesis, 
Institut f\"ur Kernphysik, Mainz, Germany 2000.
\bibitem{Kub02} G.~Kubon \emph{et al.}, Phys.~Lett.~B 
\textbf{524}, 26--32 (2002).
\bibitem{Afa01} A.~Afanasev, I.~Akushevich, N.~Merenkov, 
Phys.~Rev.~D \textbf{64}, 113009 (2001).
\bibitem{Mey94} M.~Meyerhoff \emph{et al.}, Phys.~Lett.~B \textbf{327}, 
201--207 (1994).
\bibitem{Gol01} J.~Golak \emph{et al.}, Phys.~Rev.~C \textbf{63}, 
034006 (2001).
\bibitem{Fri03} J.~Friedrich, Th.~Walcher, Eur.~Phys.~J.~A \textbf{17}, 
607--623 (2003).
\bibitem{Kim96} H.-Ch.~Kim, A.~Blotz, M.V.~Polyakov, K.~Goeke,
Phys.~Rev.~D \textbf{53}, 4013--4029 (1996).
\bibitem{Mer02} D.~Merten \emph{et al.}, 
Eur.~Phys.~J.~A \textbf{14}, 477--489 (2002).
\bibitem{Lom02} E.L.~Lomon, Phys.~Rev.~C \textbf{66}, 045501 (2002).
\bibitem{Ham04} H.-W.~Hammer, U.-G.~Mei{\ss}ner, Eur.~Phys.~J.~A 
\textbf{20}, 469--473 (2004).
\bibitem{Ma02} B.-Q.~Ma, D.~Qing, I.~Schmidt, Phys.~Rev.~C \textbf{65}, 
035205 (2002).
\bibitem{Hol02} G.~Holzwarth, hep-ph/0201138.
\bibitem{Ash04} J.D.~Ashley, D.B.~Leinweber, A.W.~Thomas, R.D.~Young, 
Eur.~Phys.~J.~A \textbf{19}, 9--14 (2004).
\end{thebibliography}
%

\end{document}